\documentclass[%
 aip,
 amsmath,amssymb,
 reprint,%
]{revtex4-1}

\usepackage{graphicx}
\usepackage{dcolumn}
\usepackage{bm}

\usepackage[utf8]{inputenc}
\usepackage[T1]{fontenc}
\usepackage{mathptmx}
\usepackage{etoolbox}

\makeatletter
\def\@email#1#2{%
 \endgroup
 \patchcmd{\titleblock@produce}
  {\frontmatter@RRAPformat}
  {\frontmatter@RRAPformat{\produce@RRAP{*#1\href{mailto:#2}{#2}}}\frontmatter@RRAPformat}
  {}{}
}%
\makeatother
\begin{document}

\preprint{AIP/123-QED}

\title{Second-order Doppler frequency shifts of trapped ions in a linear Paul trap}
\author{S. N. Miao}
\affiliation{
 State Key Laboratory of Precision Measurement Technology and Instruments, Key Laboratory of Photon Measurement and Control Technology of Ministry of Education, Department of Precision Instruments, Tsinghua University, Beijing 100084, China
}

\author{J. W. Zhang}
 \email{zhangjw@tsinghua.edu.cn}
\affiliation{
 State Key Laboratory of Precision Measurement Technology and Instruments, Key Laboratory of Photon Measurement and Control Technology of Ministry of Education, Department of Precision Instruments, Tsinghua University, Beijing 100084, China
}

\author{Y. Zheng}
\affiliation{
 State Key Laboratory of Precision Measurement Technology and Instruments, Key Laboratory of Photon Measurement and Control Technology of Ministry of Education, Department of Precision Instruments, Tsinghua University, Beijing 100084, China
}
\affiliation{
 Department of Physics, Tsinghua University, Beijing 100084, China
}

\author{H. R. Qin}
\affiliation{
 State Key Laboratory of Precision Measurement Technology and Instruments, Key Laboratory of Photon Measurement and Control Technology of Ministry of Education, Department of Precision Instruments, Tsinghua University, Beijing 100084, China
}
\affiliation{
 Department of Physics, Tsinghua University, Beijing 100084, China
}

\author{N. C. Xin}
\affiliation{
 State Key Laboratory of Precision Measurement Technology and Instruments, Key Laboratory of Photon Measurement and Control Technology of Ministry of Education, Department of Precision Instruments, Tsinghua University, Beijing 100084, China
}

\author{Y. T. Chen}
\affiliation{
 State Key Laboratory of Precision Measurement Technology and Instruments, Key Laboratory of Photon Measurement and Control Technology of Ministry of Education, Department of Precision Instruments, Tsinghua University, Beijing 100084, China
}

\author{J. Z. Han}
\affiliation{
 State Key Laboratory of Precision Measurement Technology and Instruments, Key Laboratory of Photon Measurement and Control Technology of Ministry of Education, Department of Precision Instruments, Tsinghua University, Beijing 100084, China
}

\author{L. J. Wang}
\affiliation{
 State Key Laboratory of Precision Measurement Technology and Instruments, Key Laboratory of Photon Measurement and Control Technology of Ministry of Education, Department of Precision Instruments, Tsinghua University, Beijing 100084, China
}
\affiliation{
 Department of Physics, Tsinghua University, Beijing 100084, China
}

\date{\today}

\begin{abstract}
The accurate evaluation of the second-order Doppler frequency shift (SODFS) of trapped ions in a linear Paul trap has been studied with experiments and molecular dynamics (MD) simulations. The motion of trapped ions in the trap has three  contributions, and we focus on the ion excess micromotion, which is rarely discussed when evaluating the SODFS. Based on the hypothesis that the ion density is uniformly distributed in the radial direction, we propose a new model to accurately evaluate the total SODFS for ion microwave clocks. The effectiveness of the model has been verified both in simulation and experiment, especially for ion ensemble with temperature less than 100 mK. We believe that our new model offers advantages in accurately evaluating the SODFS for the ion trap, especially those of laser-cooled ion microwave clocks based on large ion clouds.
\end{abstract}

\maketitle


\section{\label{sec:level1}Introduction}
Several recent studies have focused on developing trapped-ion microwave clocks because of the long interaction time between the ions and the applied radiation field \cite{wineland1990progress,berkeland1998laser,tjoelker2002mercury,burt2021demonstration,park2007171yb+,phoonthong2014determination,mulholland2019laser,xin2022laser,miao2015high,miao2021precision,qin2022high}. For instance, for several ${ }^{199} \mathrm{Hg}^{+}$ ions confined in a linear radio frequency (RF) Paul trap and cooled to the Doppler limit, the fractional second-order Doppler frequency shift (SODFS) may be as low as $2\times10^{-18}$ \cite{wineland1990progress}. However, when a large number of ions are confined in the trap, most ions are not at the nodal line of the trap’s RF electric field. This displacement of the ions gives rise to an excess micromotion and consequently a large frequency shift. To construct trapped-ion microwave clocks of high accuracy, the SODFS from ion micromotion must be carefully evaluated. 

According to Ref. \onlinecite{berkeland1998minimization}, the motion of a single ion of mass $M$ and charge $Q$ in a linear Paul trap has three contributions: a secular motion of large amplitude, a micromotion that oscillates at the RF frequency $\Omega$ and an excess micromotion driven by the RF field due to deviation from the nodal line of the RF field. In general, the energy of the first two parts is approximately equal. In Ref. \onlinecite{prestage1999higher}, the total fractional SODFS of trapped ions in a linear Paul trap is expressed as
\begin{equation}
\label{equ01}
\frac{\Delta f}{f}=-\frac{3 k_{B} T}{2 M c^{2}}\left(1+\frac{2}{3} N_{d}^{k}\right),
\end{equation}
where $k_B$ denotes the Boltzmann constant, $T$ the kinetic temperature characterizing the secular motion of the ions, $c$ the vacuum speed of the light, and ${N_d}^k$ a parameter related to the configuration of the trap. In the linear quadrupole trap with only a small number of non-interacting ions, space charge effects are negligible, with ${N_d}^k$ being approximately 1. This is a consequence of the equality of the average secular energy and average micromotion energy in a harmonic quadrupole trap and the lack of micromotion in the axial direction. For large ion clouds, the space charge interaction grows larger and cannot be ignored, and the value of ${N_d}^k$ is typically 3. However, the contribution of excess micrmotion of trapped ions to the SODFS is not taken into account in Eq. (\ref{equ01}), which is necessary to accurately evaluate the SODFS for microwave atomic clocks, especially those of laser-cooled ion microwave clocks based on large ion clouds.

In this paper, we focus on the excess micromotion of trapped ions in the linear Paul trap. Based on the hypothesis that the ion density is uniformly distributed in the radial direction, we propose a new model to accurately evaluate the total fractional SODFS for ion microwave clocks. The effectiveness of the model has been verified both in simulation and experiment, especially for ion ensemble with temperature less than 100 mK.

\section{Experimental setup}
\begin{figure}[t!]
\centering
\includegraphics[width=8cm]{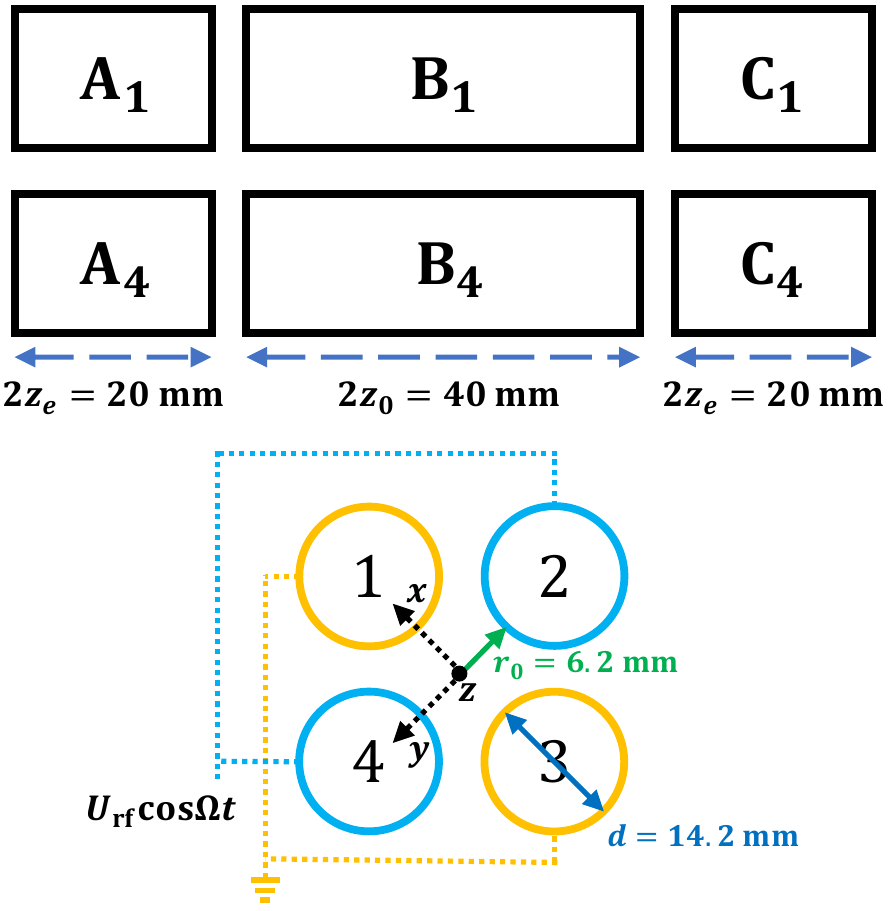}
\caption{\label{fig:fig_01} (Color online) Schematic of our linear quadrupole Paul trap. The rod diameter ($d$) and the inner radius ($r_0$) are 14.2 and 6.2 mm, respectively. The central trap region ($2z_0$) is 40 mm. The origin of the three-dimensional (3D) coordinate axis coincides with the geometric center of the ion trap.}
\end{figure}

The experimental setup has been described in more detail elsewhere \cite{miao2021research,xin2021research}; here, a brief description suffices (Fig. \ref{fig:fig_01}). It consists of four rod electrodes (1,2,3,4) with a diameter of $d=14.2$ mm, and each rod is segmented into three parts (A,B,C). The minimum distance between the nodal line of the trap and the electrode surfaces is $r_0=6.2$ mm. The lengths of the trapping part (B) and the remaining parts (A,C) of each rod are $2z_0=40$ mm and $2z_e=20$ mm, respectively. Confinement of ions is achieved by applying a radio frequency (RF) voltage $U_{\mathrm{rf}} \cos (\Omega t)$ to one pair of diagonal electrode rods ($\mathrm{B}_2$,$\mathrm{B}_4$) and a DC voltage $U_{\mathrm{end}}$ to end electrodes (A,C).

\begin{figure}[t!]
\centering
\includegraphics[width=8.5cm]{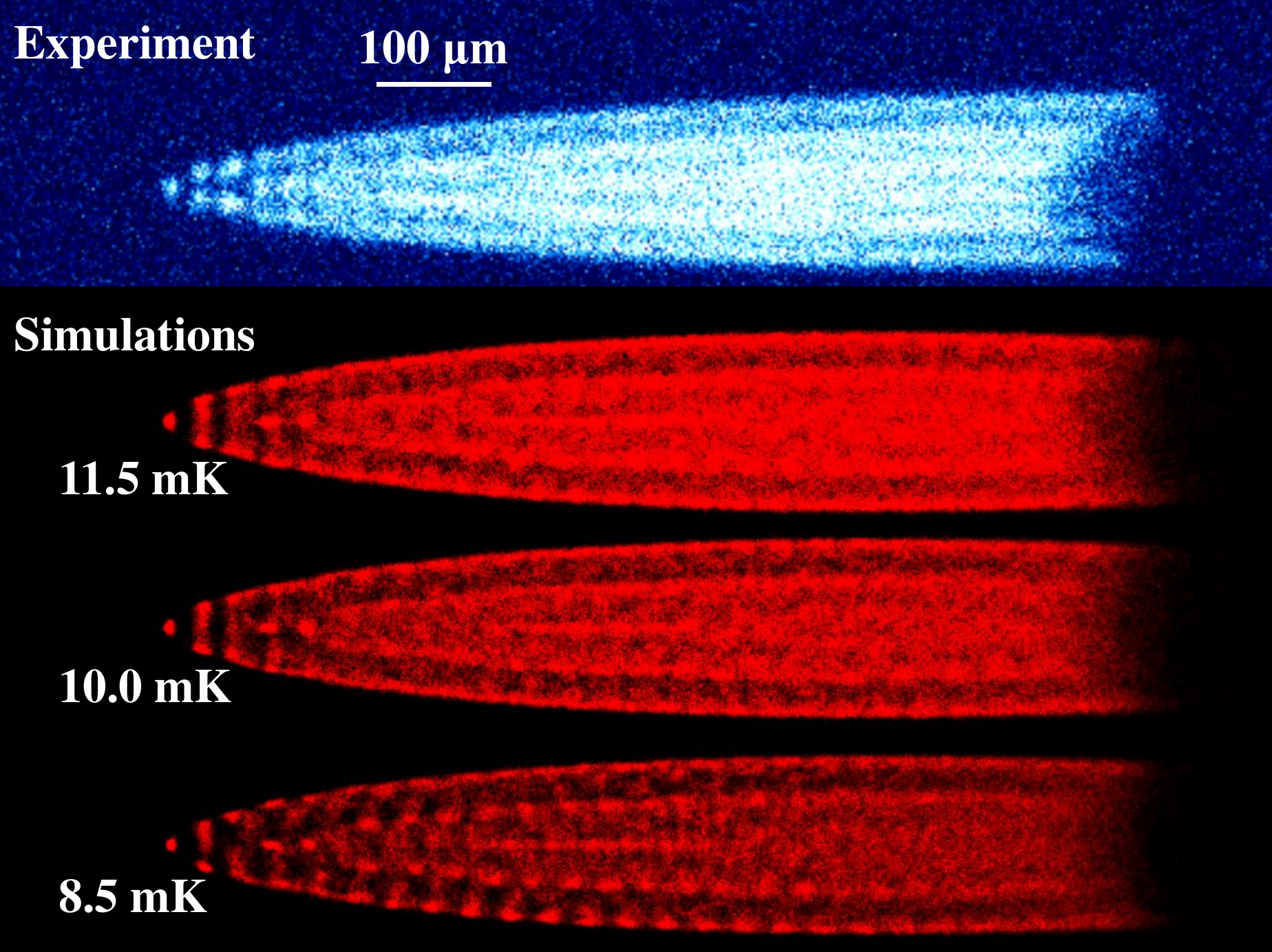}
\caption{\label{fig:fig_02} (Color online) Temperature determination of laser-cooled ${ }^{174} \mathrm{Yb}^{+}$ ions. The crystal contains about 406 ions and the CCD exposure time is 2 s. A series of simulated ion crystals at different temperatures are compared to the experimental CCD image. The best agreement is achieved for a temperature of about 10.0 mK.}
\end{figure}

With two diode lasers ($\lambda = 369$ and 935 nm), a typical Coulomb crystal containing about 406 ${ }^{174} \mathrm{Yb}^{+}$ ions is observed by a cooled CCD camera (Fig. \ref{fig:fig_02}). In order to extract the ion temperature from the experimentally produced crystal, some simulations were performed by the molecular dynamics (MD) approach. The best agreement is achieved for temperature of about 10.0 mK by comparing the experimentally observed CCD image to simulated ones. However, the total energy of ions ($3/2k_B \cdot$ 4.2 K) is much higher than that of secular motion ($3/2k_B \cdot$ 10.0 mK), which is attributed to the excess micromotion of trapped ions. Therefore, the energy of ion excess micromotion is dominant for large Coulomb crystals and is essential in the evaluation of the SODFS.

\section{Excess micromotion in a linear Paul trap}
\begin{figure}[t!]
\centering
\includegraphics[width=8cm]{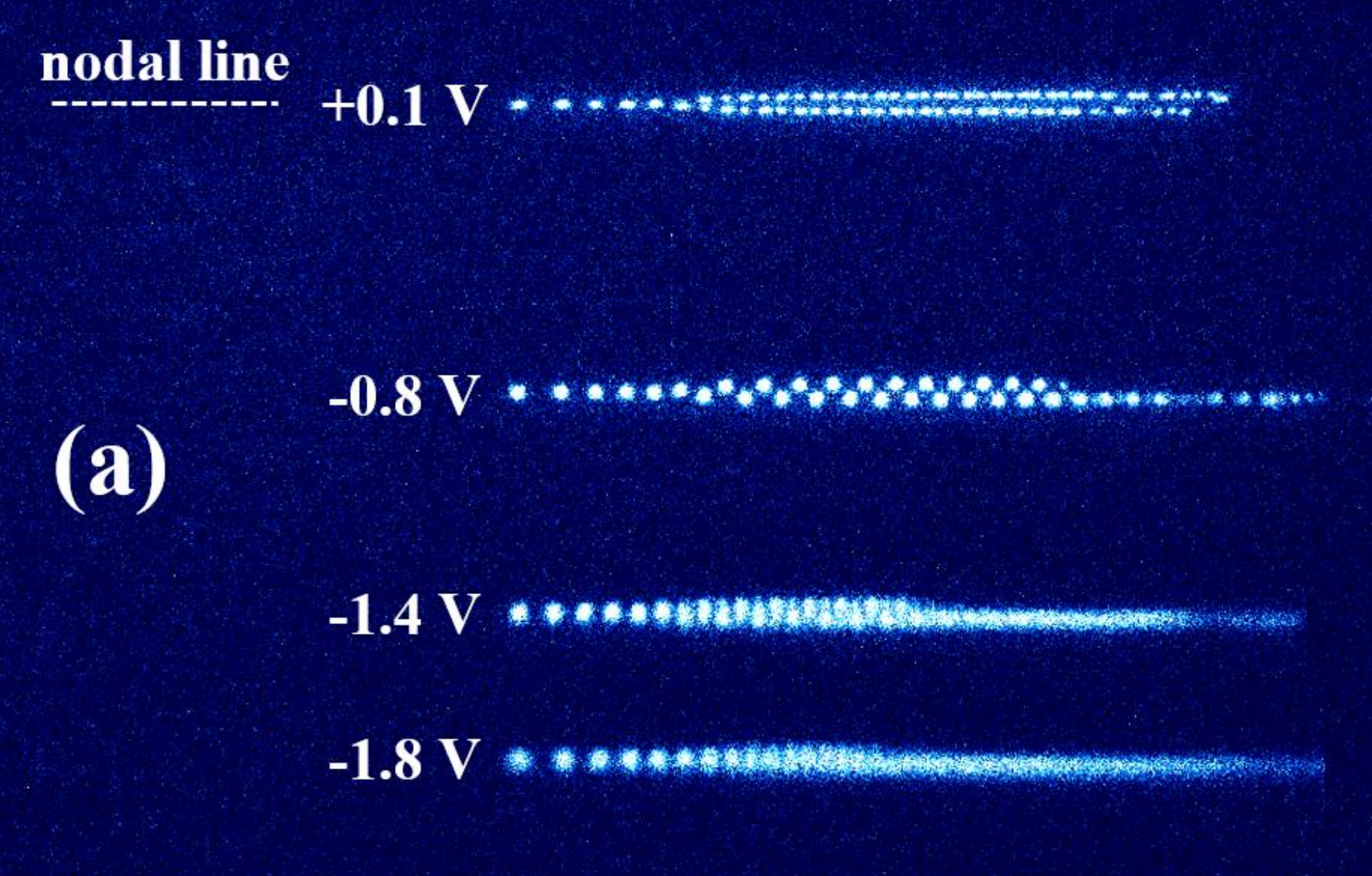}\\
\includegraphics[width=8.5cm]{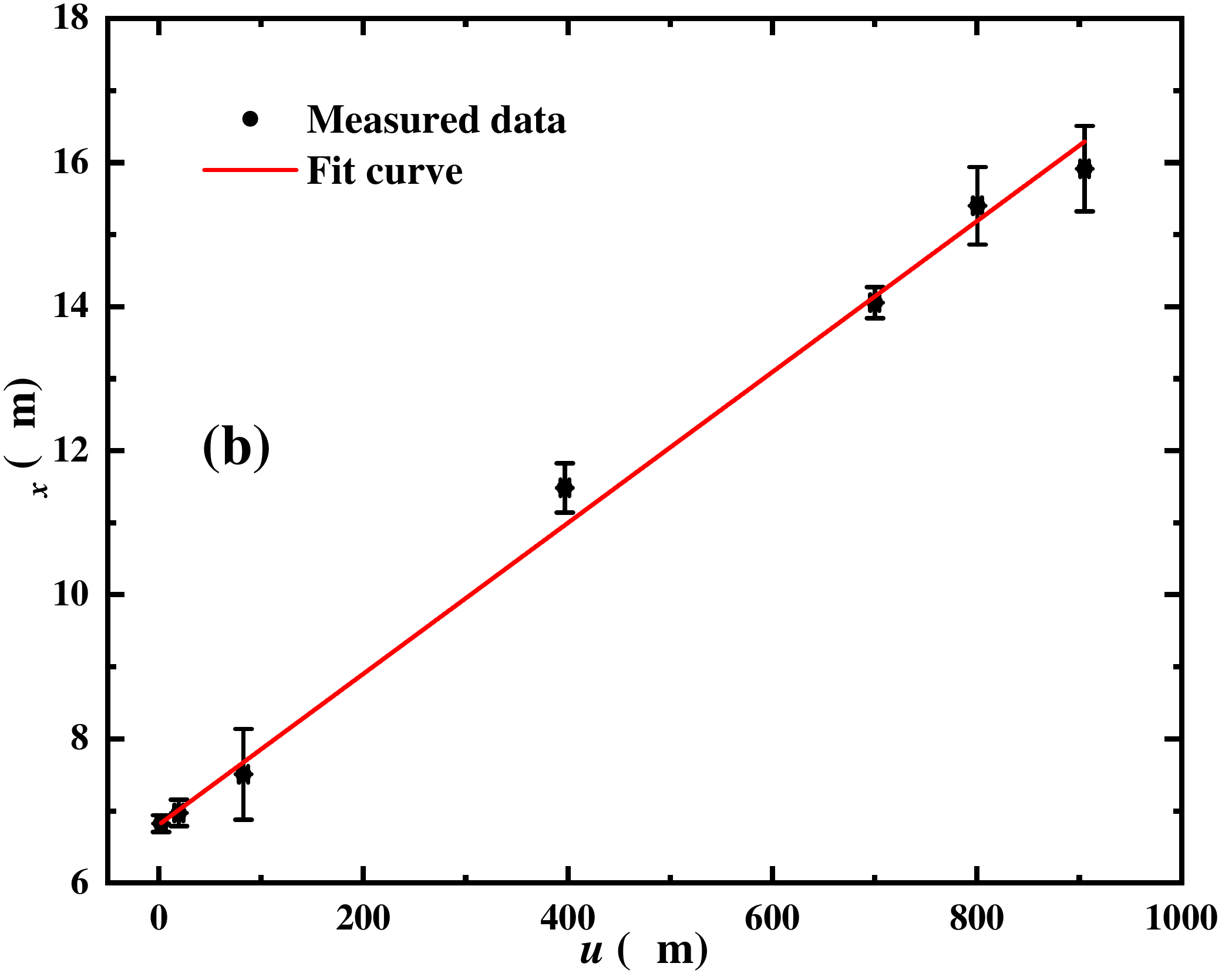}
\caption{\label{fig:fig_03} (Color online) (a) The same group of ${ }^{174} \mathrm{Yb}^{+}$ ions under different compensation voltages. The optimal value is 0.1 V, corresponding to the minimization of excess micromotion. (b) Relationship between the amplitude of ion excess micromotion in the $x$ direction and the average distance of ions from the nodal line of the RF field. The dots and solid line signify raw experimental data and theoretical curve, respectively. Each point is the average of the leftmost five ions from the CCD camera.}
\end{figure}

As is well known, when the ion deviates from the nodal line of the RF field, it experiences an excess micromotion of amplitude $q u/2$, where $u$ denotes the average distance of the ion from the nodal line, and $q$ a dimensionless parameter defined as 
\begin{equation}
\label{equ03}
q=\frac{2 Q U_{\mathrm{rf}}}{M r_{0}^{2} \Omega^{2}}.
\end{equation}
In our experiment, the ions deviate from the nodal line of the RF field by adjusting the compensation voltage (Fig. \ref{fig:fig_03}(a)). The optimal compensation voltage is 0.1 V, corresponding to the minimization of excess micromotion.

The average distance of the ion from the nodel line ($u$) is determined by the image shift on the camera and the magnification of the imaging system. The amplitudes of individual ions in the $x$ direction ($\sigma_x$) can be obtained by fitting the image from the CCD camera (see Appendix for details). In order to reduce the uncertainty of the results, only the five clearest ions on the far left are considered (Fig. \ref{fig:fig_03}(a)). Each point in Fig. \ref{fig:fig_03}(b) is the average of five ions. The results show that the amplitude of ion excess micromotion increases linearly with the average displacement from the nodal line, which is consistent with the theory in Ref. \onlinecite{berkeland1998minimization}.

\section{A new model to evaluate the SODFS of ions}

According to the above conclusions about ion excess micromotion, the average kinetic energy of excess micromotion for the $i$-th ion in a large ion cloud is expressed as
\begin{equation}
\label{equ03}
E_{K i} = \frac{1}{2} M \Omega^{2}\left(\frac{\sigma_{i}}{\sqrt{2}}\right)^{2}=\frac{1}{4} M \Omega^{2} \sigma_{i}^{2}=\frac{1}{16} M \Omega^{2} q^{2} u_{i}^{2},
\end{equation}
where $\sigma_i$ denotes the amplitude of ion excess micromotion. Therefore, the total fractional SODFS for the trapped ions is given by
\begin{equation}
\label{equ04}
\frac{\Delta f}{f}=-\frac{3 k_{B} T}{2 M c^{2}}\left(1+\frac{2}{3} N_{d}^{k}\right)-\frac{q^{2} \Omega^{2} \left\langle u^{2} \right\rangle}{16c^{2}},
\end{equation}
where $\langle\cdots\rangle$ denotes the average of all ions. The three parts in the formula represent the three motions of trapped ions in the quadrupole trap, respectively. The first two parts are determined by the secular temperature $T$ of the ion ensemble, and ${N_d}^k = 1$ due to the equality of the average secular energy and average micromotion energy. In MD simulation, the secular temperature of ions is expressible as
\begin{equation}
\label{equ05}
T=\frac{1}{3 N k_{B}} M \sum_{i}^{N}\left\langle\bar{v}_{i}^{2}\right\rangle,
\end{equation}
where $N$ denotes the total ion number, $\bar{v}_{i}$ the secular velocity of the $i$-th ion defined by averaging over one RF period, and $\langle\cdots\rangle$ the average over many RF periods. In our experiment, the secular temperature can be obtained by measuring the Gaussian broadening of the spectral linewidth \cite{zuo2019direct,han2021toward,miao2022sympathetic}. The third part in Eq. (\ref{equ04}) is contributed by the ion excess micromotion, which is related to the number of ions and electrical parameters. In the following, we focus on analyzing the third part.

\begin{figure}[t!]
\centering
\includegraphics[width=8.5cm]{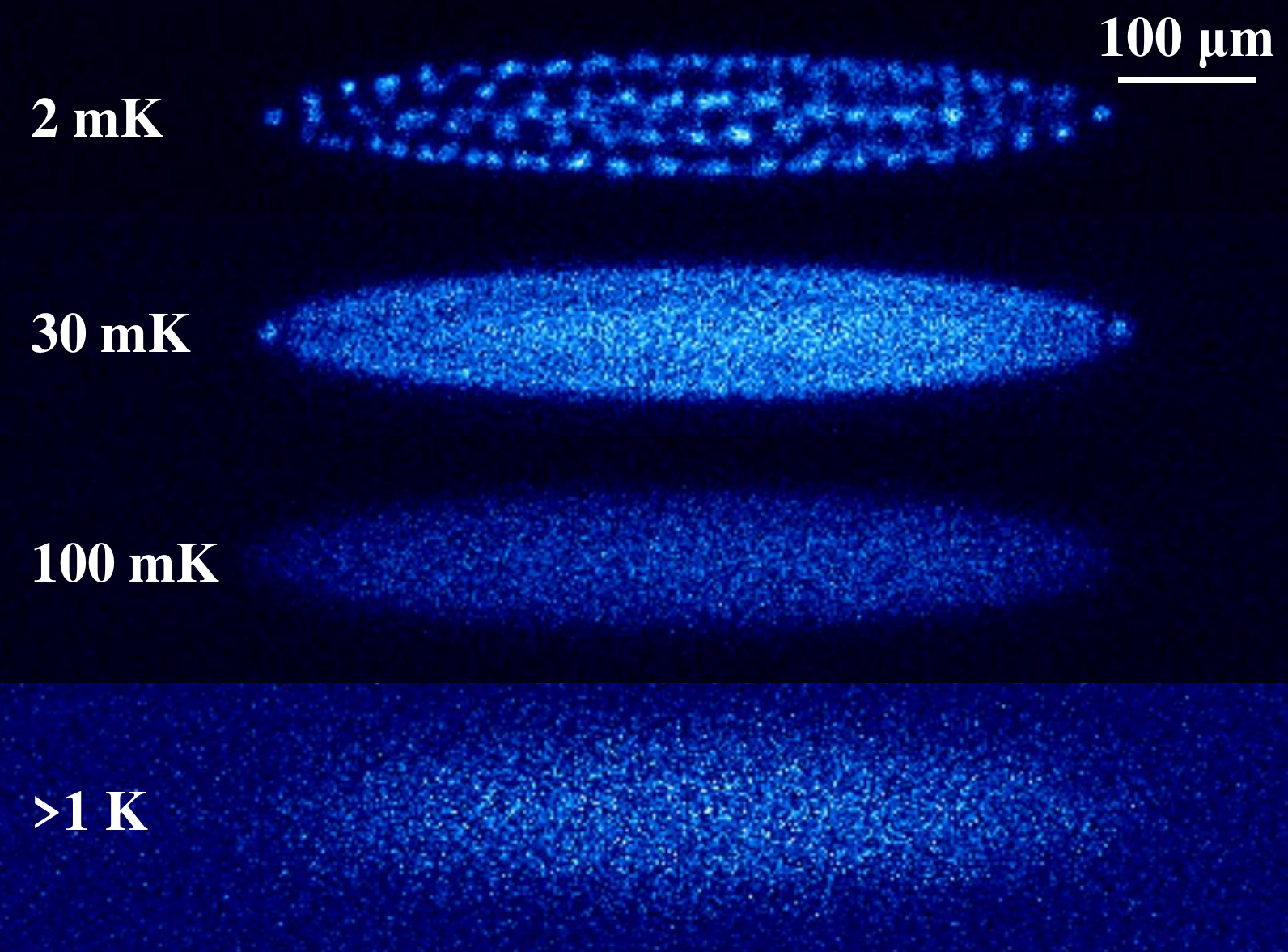}
\caption{\label{fig:fig_04} (Color online) A series of experimental pictures of the same group of ${ }^{174} \mathrm{Yb}^{+}$ ions at different temperatures under the influence of fixed electrical parameters. The CCD camera exposure time is 2 s. The volume of the ion ensemble is basically unchanged for $T \textless 100 $ mK.}
\end{figure}

The previous theories for calculating the SODFS of large ion clouds in multipole traps are based on the hypothesis that the radial density of ions obeys the Boltzmann distribution \cite{prestage1999higher,fisk1997trapped}. However, this hypothesis is only valid for large ion clouds at high temperature, but not for low-temperature Coulomb crystals. In an ion Coulomb crystal, the ions are distributed in multiple so-called ellipsoidal shells, which is obviously inconsistent with the Boltzmann distribution. Therefore, based on the zero temperature charged liquids model \cite{hornekaer2000single,o1981centrifugal,wineland1987ion,hornekaer2001structural}, we propose that the ion density is uniformly distributed in the radial direction. In Fig. \ref{fig:fig_04}, we measured the volume of the same ion ensemble at different temperatures under the influence of fixed electrical parameters. It is clear that the volume of the ion ensemble is basically unchanged for $T \textless 100 $ mK, which shows that the model is applicable to ion ensembles in non gaseous state. 

Under the premise of the above hypothesis, Eq. (\ref{equ04}) becomes
\begin{equation}
\label{equ06}
\frac{\Delta f}{f}=-\frac{3 k_{B} T}{2 M c^{2}}\left(1+\frac{2}{3} N_{d}^{k}\right)-\frac{q^{2} \Omega^{2}  u_{\mathrm{eff}}^{2}}{16c^{2}},
\end{equation}
where $u_{\mathrm{eff}}$ denotes the equivalent distance of all ions from the nodal line of the RF field. In an analogous calculation to the moment of inertia of an ellipsoid, the equivalent distance is determined to be $u_{\mathrm{eff}} = \sqrt{10}/{5} R$, where $R$ denotes the radial size of the spheroid fit to the outer ion cloud envelope. Considering that the measurement uncertainty of the half length ($L$) of the outer ion cloud envelope is relatively smaller than that of $R$, Eq. (\ref{equ04}) can also be expressed as
\begin{equation}
\label{equ07}
\frac{\Delta f}{f}=-\frac{3 k_{B} T}{2 M c^{2}}\left(1+\frac{2}{3} N_{d}^{k}\right)-\frac{3 Q^{2}}{40\pi\epsilon_0 M c^2} \cdot \frac{N}{L},
\end{equation}
where $\epsilon_0$ denotes the permittivity of vacuum.

\begin{figure}[t!]
\centering
\includegraphics[width=8cm]{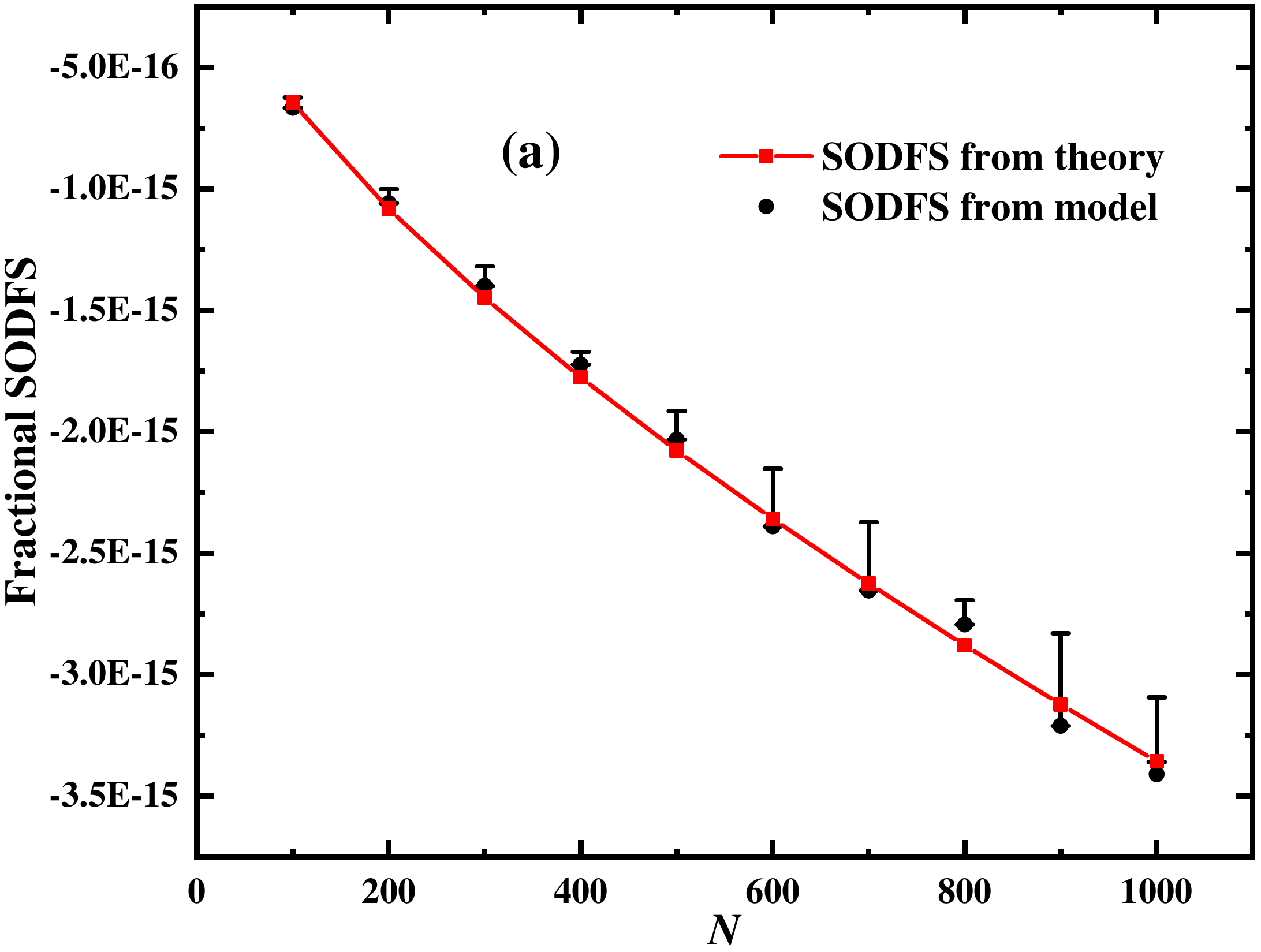}\\
\includegraphics[width=8cm]{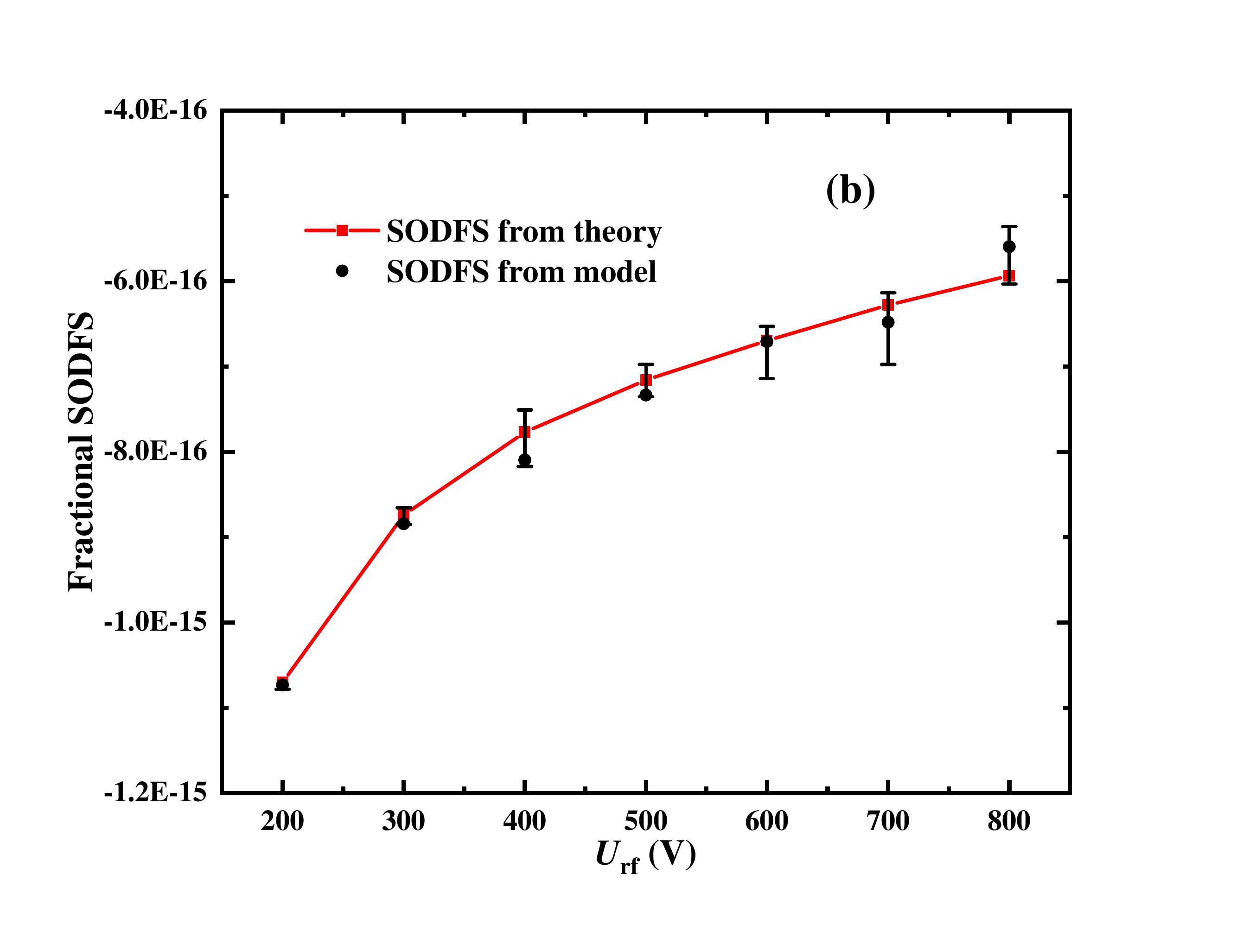}\\
\includegraphics[width=8cm]{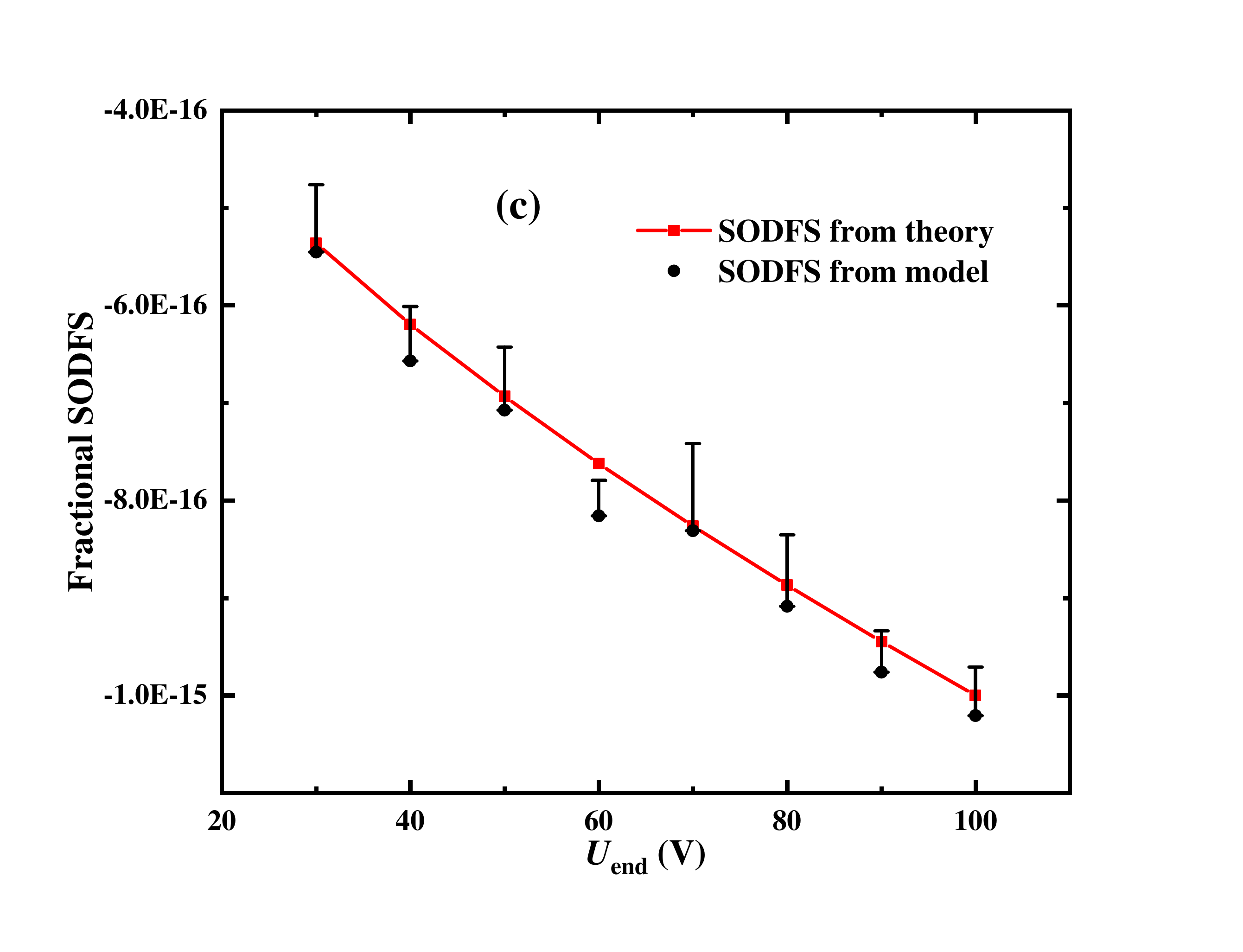}
\caption{\label{fig:fig_05} (Color online) A series of MD simulations performed to verify the effectiveness of the model in Eq. (\ref{equ06}). During the simulation, the secular temperature of ${ }^{174} \mathrm{Yb}^{+}$ ions in equilibrium is controlled at 10 mK. (a) Determination of the relationship between the total fractional SODFS of the trapped ions and the number of ions under the influence of fixed electrical parameters ($U_{\mathrm{rf}} = 400$ V and $U_{\mathrm{end}} = 60$ V). Determination of the relationship between the total fractional SODFS of the trapped ions and (b) RF voltage $U_{\mathrm{rf}}$, (c) DC voltage $U_{\mathrm{end}}$. Increasing $U_{\mathrm{end}}$ or decreasing $U_{\mathrm{rf}}$ will increase the total fractional SODFS of the trapped ions.}
\end{figure} 

A series of MD simulations were performed to verify the effectiveness of our new model. The advantage of MD simulation is that the velocities of all ions can be extracted, from which the total fractional SODFS of the trapped ions can be directly calculated using $\Delta f / f=-\left\langle v^{2}\right\rangle /\left(2 c^{2}\right)$ (SODFS from theory). On the other hand, the total SODFS is evaluated with our new model (SODFS from model).
During the simulation, the secular temperature of ions in equilibrium is controlled at 10 mK by coupling all ions to a Langevin bath \cite{cortes1985generalized,ford1987quantum,attal2007langevin}. In simulations to determine the relationship between the total fractional SODFS of ions and $N$, we set $U_{\mathrm{rf}} = 400$ V and $U_{\mathrm{end}} = 60$ V. Plot (Fig. \ref{fig:fig_05}(a)) shows the total fractional SODFS increases with the number of ions, which is a consequence of the expanding size of ions. The results from the new model are consistent with those calculated by the theoretical formula.
In Fig. \ref{fig:fig_05}(b) and \ref{fig:fig_05}(c), the relationship between the total fractional SODFS of ions and electrical parameters is shown. It can be seen that the SODFS of the trapped ions evaluated by the above two methods are consistent. Increasing $U_{\mathrm{end}}$ or decreasing $U_{\mathrm{rf}}$ will increase the radial size of the ion ensemble, and then increase the total SODFS of ions.

Apart from that, there are some experimental results that can be well explained by our new model in Eq. (\ref{equ06}). In Ref. \onlinecite{miao2021precision}, the 0-0 ground-state hyperfine transition frequency of ${ }^{113} \mathrm{Cd}^{+}$ is consistent with previously reported values when evaluating the total SODFS of ions using the model in Eq. (\ref{equ06}), but not using the previous theory in Eq. (\ref{equ01}). For the ${ }^{171} \mathrm{Yb}^{+}$ ion Coulomb crystal in Ref. \onlinecite{mulholland2019laser}, the radial radius of the spheroid fit to the crystal envelope is $R = 80(5)$ $\mu$m, and the secular temperature of the crystal is lower than 50 mK. Therefore, the total fractional SODFS for the ytterbium-ion microwave frequency standard given by our new model is $-1.30(16)\times10^{-14}$, which is consistent with the result $\left(\textless -2.0(0.5)\times10^{-14}\right)$ in Ref. \onlinecite{mulholland2019laser}.

The effectiveness of our new model has been verified both in simulation and experiment. We note in particular that the excess micromotion of ions is important when evaluating the total SODFS of the traped ions, especially for low-temperature Coulomb crystals. In addition, both the number of ions and electrical parameters affect the energy of ion excess micromotion, which in turn affects the total SODFS of ion microwave clocks.

\section{Conclusion}
In summary, the Second-order Doppler frequency shift of the trapped ions in a linear Paul trap has been studied in detail. The motion of ions in the trap has three contributions, and we focused on the ion excess micromotion. Based on the hypothesis that the ion density is uniformly distributed in the radial direction, we propose a new model to evaluate the total SODFS for ion microwave clocks. The effectiveness of the model has been verified both in simulation and experiment, especially for ion ensemble with temperature less than 100 mK. According to the model, ion temperature, ion number and electrical parameters should be taken into account in order to reduce the total SODFS of the trapped ions. The model and the analytical results would be very useful to experimental physicists advancing microwave atomic clock technology relying on fractional SODFS evaluations.

\begin{acknowledgments}
The authors thank Z. B. Wang, K. Miao, C. F. Wu, H. X. Hu, W. X. Shi and T. G. Zhao for their helpful assistance and discussions. This work is supported by National Natural Science Foundation of China (12073015), Tsinghua University Initiative Scientific Research Program, and Beijing Natural Science Foundation (1202011).
\end{acknowledgments}

\section*{Data Availability Statement}
The data that support the findings of this study are available from the corresponding author upon reasonable request.

\appendix*

\section{Measuring the amplitudes of trapped ions by imaging}

Referring to Ref. \onlinecite{srivathsan2019measuring}, the image recorded on the CCD camera is a convolution of the imaging point-spread function (PSF) and the ‘true image’ of the ion. Assuming both the PSF and the true image to be Gaussian spots, the width of the recorded image can be approximated as
\begin{equation}
\sigma^{2}=\sigma_{\mathrm{PSF}}^{2}+M^{2} \sigma_{i}^{2},
\end{equation}
where $M$ is the magnification of the imaging system, $\sigma_{\mathrm{PSF}}$ the width of the imaging PSF caused by the diffraction of the system, and $\sigma_{i}$ the desired amplitude of the ions. In our system, the magnification is determined to be $M = 5.96$. $\sigma_{\mathrm{PSF}}$ is determined to be 2.38 $\mu$m, which depends on the parameters of the camera (PF10545MF-UV) and the optical diffraction limit.

\bibliography{aipsamp}

\end{document}